\documentstyle[seceq]{ptptex}

\newcommand{\bphi}{{\hbox{\boldmath $\phi$}}}
\newcommand{\bchi}{{\hbox{\boldmath $\chi$}}}

\newcommand{\bpsi}{{\hbox{\boldmath $\psi$}}}

\newcommand{\calh}{{\cal H}}
\newcommand{\calR}{{\cal R}}
\newcommand{\ksg}{k \sigma_g}
\preprintnumber{
OU-TAP 72}
\title{
Super-Horizon Scale Dynamics\\ of Multi-Scalar Inflation
}

\author{
Misao {\sc Sasaki}\footnote{Electronic address: 
misao@vega.ess.sci.osaka-u.ac.jp}
and
Takahiro {\sc Tanaka}\footnote{Electronic address: 
tama@vega.ess.sci.osaka-u.ac.jp} 
}

\inst{
Department of Earth and Space Science, Graduate School of Science,\\
Osaka University, Toyonaka 560-0043
}

\recdate{
}

\abst{
We consider the dynamics of a multi-component scalar field 
on super-horizon scales in the context of inflationary cosmology. 
We present a method to solve the perturbation equations on super-horizon
scales, i.e., in the long wavelength limit,
by using only the knowledge of spatially homogeneous background 
solutions. In doing so, we clarify the relation between the 
perturbation equations in the long wavelength limit
and the background equations. Then as a natural extension of our
formalism, we provide a strategy to study super-horizon scale
perturbations beyond the standard linear perturbation theory.
Namely we reformulate our method so as to take into account
the nonlinear dynamics of the scalar field.
}

\begin{document}

\maketitle

\small
\section{Introduction}

Study of multi-scalar inflation is a current topic 
in cosmology, motivated mainly by 
the fact that supergravity theories suggest the existence of 
many flat directions in the scalar field potential\cite{super}. 
Observationally we will be able to determine the accurate spectrum of 
primordial density fluctuations in the near future, e.g., by
the next-generation MAP and PLANCK satellites\cite{MapPlanck}. 
At that time, to test a variety of models of inflation, we will need a
systematic method to evaluate the spectrum for a wide class of inflaton
models.

In this paper, we consider perturbations of a 
multi-component scalar field in the long wavelength limit. That is, we 
consider modes whose wavelengths exceed the Hubble horizon scale. 
In the case of single-scalar inflation, 
the evolution of a perturbation in the long wavelength limit
is very simple. The behavior of the adiabatic
growing mode is specified by $\calR_c=$ constant, where 
$\calR_c$ is the spatial curvature perturbation of
 the comoving hypersurface. 
In the case of multi-scalar inflation, the same is not true. 
Even if the wavelength of a perturbation exceeds the horizon scale,
 $\calR_c$ changes in time. 
The evolution of super-horizon scale perturbations for 
multi-scalar inflation has been investigated in some particular models 
that allow analytical treatments\cite{parti}. However, since
the feasibility of analysis of a model does not imply the 
viability of the model, it is necessary to develop a method that has
a wide range of applicability. In this context, a rather general 
framework to study this issue has been given by 
Sasaki and Stewart\cite{SasSte} under the assumption of 
the slow rolling evolution of a multi-component scalar field. 
However, it may well happen that some components of the scalar field 
do not satisfy the slow rolling condition during inflation. 
Furthermore, the slow rolling condition will
be eventually violated toward the end of inflation when the reheating
(or preheating) commences. Hence it is much more desirable to exclude
the slow rolling assumption. In this connection, Taruya and Nambu 
have recently discussed a method to obtain the general solution for 
long wavelength perturbations\cite{TarNam}. Unfortunately, however,
they have not clarified several delicate issues associated with
super-horizon scale perturbations that are characteristic of general
relativity. Here we develop a general framework to study super-horizon
scale perturbations without assuming slow rolling. Note that, although
we use the terminology of `multi-scalar inflation', our framework
will be valid for any stage of the universe as long as the energy
momentum tensor is dominated by a multi-component scalar field.
Then we extend our formalism so as to include the effect of nonlinearity
of the scalar field potential. 

Suppose we are given a set of equations that governs the evolution of 
a system. If we know a complete set of solutions to 
this set of equations, we have no difficulty in constructing 
a general solution of the perturbation around a fixed background
 solution. By definition, a complete set of solutions contains 
 a sufficient number of constants of motion that parametrize the 
different solutions. If we take a derivative with respect to one of 
such constants of motion, a solution to the perturbation equations 
can be obtained. The same is true for the cosmological perturbations.
If we could obtain a full set of solutions to the Einstein equations 
coupled with the equations of matter fields, it would be 
trivially easy to spell out the perturbation around a fixed 
background universe. But it is hopeless to expect so.
Nevertheless, if we restrict our attention to spatially homogeneous
and isotropic universes, we may be able to obtain a full set of 
solutions. Then one may expect that the perturbation in the long
wavelength limit is obtained from the knowledge of this 
restricted class of solutions. In fact, although restricted by
the slow rolling assumption, the formula for the curvature perturbation
on the comoving hypersurface derived in Ref.~\citen{SasSte} is a good 
example of such a case.

Let us consider an $n$-component scalar field 
whose action is given by 
\begin{equation} 
 S_{matter}=-\int d^4 x{\sqrt{-g}\over 2} 
   \left(g^{\mu\nu}\bphi_{,\mu}\cdot\bphi_{,\nu}+ U(\bphi)\right),
\end{equation}
where we have used vector notation to represent the $n$-component
scalar field. That is, $\bphi\cdot\bphi=h_{pq}\phi^p\phi^q$,
where $h_{pq}$ is the metric in the scalar field space. 
For simplicity we assume $h_{pq}=\delta_{pq}$ in the discussions 
below, but the generalization to a non-trivial $h_{pq}$ is 
straightforward, which we shall discuss later.
The equations for the spatially flat, homogeneous and isotropic
configuration are 
given by 
\begin{eqnarray}
&& {\ddot\phi^p}+3H {\dot\phi^p}+ U^{|p}=0, 
\label{bgcos}
\\
&& H^2={1\over 3}\left({1\over 2}{\dot\bphi}^2+U\right), 
\label{frcos}
\end{eqnarray}
where dot means a derivative with respect to the cosmological time $t$.
We adopt the units $8\pi G=c=1$, $H:=\dot a/a$ is the Hubble parameter
with $a$ being the cosmic scale factor, and 
$U^{|p}=h^{pq}\partial U/\partial\phi^q$.
A solution to these equations gives a background solution in the 
context of cosmological perturbation theory. 
Hence we refer to these equations as the background equations. 

Substituting Eq.~(\ref{frcos}) into Eq.~(\ref{bgcos}), we have 
$n$ coupled second order differential equations. 
Now, suppose that we know a complete set of solutions 
for these equations. 
The complete set of solutions will contain $2n$ 
parameters, $\lambda^{\alpha}$ ($\alpha=1,2,\cdots,2n$)
 which distinguish the various solutions. 
One of them, say $\lambda^1$, just represents the change in 
the origin of the time coordinate, $t_0$. 
So let us denote the general solution by 
$\bphi(t+\lambda^1,\lambda^{a})$, where $a$ runs from $2$ to $2n$. 
Then a naive anticipation is that we would obtain the $2n$ solutions 
for the perturbation in the long wavelength limit by 
$\partial\bphi(t+\lambda^1,\lambda^{a})/\partial\lambda^{\alpha}$,
or more concisely by 
$\partial\bphi(t,\lambda^a)/\partial\lambda^{\alpha}$
where $\lambda^1=t$. 

However, things are not so simple. 
First of all, it is not guaranteed if we can restrict our 
consideration to spatially homogeneous and isotropic configurations  
from the beginning to obtain the perturbation in the long wavelength
 limit. Furthermore, it is well known that, when dealing with 
cosmological perturbations particularly on super-horizon scales,
 it is essential to make a clear statement about the choice of gauge
before one talks about the behavior of the perturbation.
In the above discussion, it was not clarified in what gauge 
the solutions were given, even if they actually would describe
the long wavelength perturbation. One of the purposes of this paper
is to clarify these points.

The paper is organized as follows. In \S 2, we consider the linear 
perturbation equations in the long wavelength limit and clarify its
relation to the background equations. 
In \S\S 2.1, we introduce the basic notation and list the 
basic equations for the perturbation of a multi-component scalar field. 
In \S\S 2.2, we discuss the long wavelength perturbation 
in the $B=\dot H_L=0$ gauge, where $B$ is the shift vector 
perturbation and $H_L$ is the trace part of the spatial metric
 perturbation (see Eq.~(\ref{metric}) below).  
In this gauge we find the $e$-folding number $N:=\log a$ is 
unperturbed even under the presence of the perturbation, hence it
 can be regarded as the natural time coordinate instead of the 
cosmological time, $t$. Then we show that there exists 
a simple relation between the perturbation equations in the 
long wavelength limit and the derivative of the background 
equations with respect to the parameters
$\{\lambda^{\alpha}\}:=\{N,\lambda^a\}$. 
This result indicates there is a unique correspondence between 
the choice of the time coordinate for the background equations
and that of gauge for the perturbation equations.
To strengthen our assertion, in Appendix we consider the 
perturbation equations in a couple of other choices of time coordinates
and show that this is indeed the case.
Then using this relation, we give a formula to calculate the scalar 
field perturbation in the flat slicing from the solutions of 
the background equations. Very recently the same formula has been 
obtained independently by Kodama and Hamazaki\cite{KodHam} by a
different approach.
In \S\S 2.3, we use this formula to evaluate 
the curvature perturbation on the comoving hypersurface at the 
end of inflation. Then we take the slow
rolling limit and recover the result of Sasaki and Stewart\cite{SasSte}.
In \S 3, we consider an extension of our formalism beyond 
the limitation of the standard linear perturbation theory.
 There we only assume the smallness of the spatial metric 
perturbation and the non scalar-type perturbation but the other 
components of perturbations are not supposed to be small. 
We find almost everything goes parallel to the linear case. 
The summary of this paper is given in \S 4.

Throughout this paper, we follow the notation and sign convention 
used in Kodama and Sasaki\cite{KodSas}. 

\section{Linear Perturbation}
\subsection{Basic equations}

First we write down a set of well known equations for 
cosmological perturbations.
We consider only the scalar-type metric perturbation
and write the perturbed metric as 
\begin{equation}
ds^2=a^2\left[-(1+2AY) d\eta^2-2B\,Y_j\,d\eta dx^j
+\bigl((1+2H_LY)\delta_{ij}
  +2H_T Y_{ij}\bigr) dx^i dx^j\right],
\label{metric}
\end{equation} 
where $Y$ is the spatial scalar harmonic with the eigenvalue $k^2$,
$Y_j=-k^{-1}\nabla_j Y$, and 
$Y_{ij}=k^{-2}\nabla_i\nabla_j Y+{1\over3}\delta_{ij}Y$.
We note the background equations~(\ref{bgcos}) and (\ref{frcos}) 
can be rewritten in terms of the conformal time, $\eta$, as 
\begin{eqnarray}
&& {\phi^p}{}''+2\calh{\phi^p}{}'+a^2 U^{|p}=0, 
\label{bg}
\\
&& \calh^2={1\over 3}\left({1\over 2}{\bphi'}^2+a^2 U\right), 
\label{fr}
\end{eqnarray}
where the prime represents a derivative with respect to $\eta$ and 
$\calh:=a'/a$. 

The perturbation of the scalar field equation becomes 
\begin{equation}
 {\chi^p}{}''+2\calh {\chi^p}{}'+k^2\chi^p + a^2 U^{|p}{}_{|q}\chi^q
-2\{{\phi^p}{}''+2\calh{\phi^p}{}'\} A - {\phi^p}{}' A' 
  +{\phi^p}{}'(3\calR'-\ksg)=0, 
\label{feqpert}
\end{equation}
where $\bchi=\delta\bphi$, and we have introduced $k\sigma_g:=H'_T-kB$ 
and $\calR:=H_L+{1\over 3}H_T$; the former represents the shear of
the $\eta=$ constant hypersurface and the latter the spatial curvature
perturbation. 

The $({}^0_0)$-component of the perturbed Einstein equations is
given by
\begin{equation}
 2\left[3\calh^2 A -\calh(3\calR' -\ksg) -k^2 \calR\right]
=-\bphi'\cdot\bchi'-a^2 U_{|p}\chi^p +A {\bphi'}^2.
\label{00eq}
\end{equation}
The $({}_i^0)$-component is given by
\begin{equation}
 2\left[\calh A-\calR'\right] = \bphi'\cdot\bchi.
\label{0ieq}
\end{equation}
The trace part of the $({}^i_j)$-component becomes 
\begin{equation}
 2\left(a^2 U A+\calh A'-\left[{d\over d\eta}+2\calh \right]
 \left(\calR'-{1\over 3}k\sigma_g\right)-{k^2\over 3}(A+\calR)\right)
 =\bphi'\cdot\bchi'-a^2 U_{|p}\chi^p. 
\label{ijTR}
\end{equation}
Finally, the traceless part of the $({}^i_j)$-component is
\begin{equation}
 \ksg'+2\calh \ksg -k^2(A+\calR)=0.
\label{ijTL}
\end{equation}
The long wavelength limit of the perturbation, i.e., the limit
$k^2/\calh^2\to0$, is described by
taking $k^2\to0$ in the above equations.

An important geometrical quantity which plays a central role in the
following discussion is the $e$-folding number of cosmic expansion.
The perturbed $e$-folding number of expansion is defined by 
\begin{equation}
\tilde N=\int_{\tau_0}^\tau\tilde H d\tau,
\label{defN}
\end{equation}
where $\tilde H$ is the perturbed Hubble parameter given by the 1/3 
of the expansion $\tilde\theta$ of the $\eta=$ constant hypersurface;
\begin{eqnarray}
&&\tilde H={1\over3}\tilde\theta=H(1+{\cal K}_gY);
\nonumber\\
&&{\cal K}_g:=-A+{1\over\calh}(H_L'+{1\over3}kB)
=-A+{1\over\calh}(\calR'-{1\over3}\ksg),
\label{deftilH}
\end{eqnarray}
and $\tau$ is the proper time along the integral curve of the vector
normal to the $\eta=$ constant hypersurfaces;
\begin{equation}
d\tau=(1+AY)a\,d\eta.
\end{equation}
One then readily sees that
\begin{equation}
\tilde N=\int_{\eta_0}^\eta
\left(\calh+(\calR'-{1\over3}\ksg)Y\right)d\eta.
\label{pertN}
\end{equation}
Thus the $e$-folding number will be unperturbed if we take the gauge
$H_L'=B=0$, which implies $\calR'=\ksg/3$. One can then expect
that the equations obtained by perturbing the background equations by
taking $N=\ln a$ as the time coordinate will closely resemble 
the perturbation equations in this gauge.

\subsection{$N$ as a time coordinate: the $H_L'=B=0$ gauge}
If we take $N$ as the time coordinate, the background 
equations can be written as 
\begin{eqnarray}
&& H{d\over dN}H{\phi^p_N}
  +3H^2{\phi^p_N}+U^{|p}=0, 
\label{bgN}
\\
&& H^2 \left(1-{1\over 6}{\bphi_N}^2\right)={1\over 3}U, 
\label{frN}
\end{eqnarray}
where $\displaystyle\bphi_N\equiv {d\bphi\over dN}$. 
We note that Eq.~(\ref{frN}) is the $({}^0_0)$-component
of the background Einstein equations. Together with this equation,
the trace part of the $({}^i_j)$-component of the Einstein equations
gives
\begin{equation}
{1\over H}{dH\over dN}=-{1\over 2}\bphi_N^2, 
\label{dHdN}
\end{equation}
which may be also obtained by substituting Eq.~(\ref{bgN}) into 
the $N$-derivative of Eq.~(\ref{frN}).

Let us consider the perturbation equations in the long wavelength 
limit in the $B=H_L'=0$ gauge. As we noted, in terms of the 
geometrical variables $\calR$ and $\ksg$, this gauge condition 
implies
\begin{equation}
\calR'-{1\over 3}\ksg =0. 
\label{gauge}
\end{equation}
Then Eq.~(\ref{feqpert}) reduces to 
\begin{equation}
  H{d\over dN}\left(H \chi^p_N\right)
 +3H^2\chi^p_N +U^{|p}{}_{|q}\chi^q
  +2U^{|p} A - H^2{\phi^p_N} A_N =0. 
\label{chieq}
\end{equation}
Thus among the metric variables,
the perturbed field equation contains only $A$.
We also note that $A$ in this gauge represents the perturbation of 
the Hubble parameter, $\delta H/H=-AY$,
 as seen from Eq.~(\ref{deftilH}).

{}From the $({}^0_0)$-component of the perturbed Einstein 
equations~(\ref{00eq}), one can see that 
$A$ is expressed in terms of $\bchi$ as 
\begin{equation}
 2 U A=-H^2 \bphi_N\cdot\bchi_N-U_{|p}\chi^{p},
\label{Aeq}
\end{equation}
where we have used Eq.~(\ref{fr}) to simplify the expression. 
At this point, one may substitute Eq.~(\ref{Aeq}) into 
Eq.~(\ref{chieq})
to obtain a closed second order equation for $\bchi$. 
It is to be noted that only the $({}^0_0)$-component is
necessary to derive the closed equation for $\bchi$.

Consider now the trace part of the $({}^i_j)$-component,
Eq.~(\ref{ijTR}). It reduces to 
\begin{equation}
  2A_N +{2U(\bphi)\over H^2}A =\bphi_N\cdot\bchi_N 
    -{U_{|p}\chi^p\over H^2}. 
\end{equation}
Together with Eq.~(\ref{Aeq}), this gives 
\begin{equation}
 A_N=\bphi_N\cdot\bchi_N. 
\label{An}
\end{equation}
This equation is not independent of Eqs.~(\ref{chieq}) 
and (\ref{Aeq}) because of the contracted Bianchi identities.
In fact, Eq.~(\ref{An}) can be directly verified 
by taking the $N$-derivative of Eq.~({\ref{Aeq}) 
and using Eqs.~(\ref{bgN}), (\ref{frN}) and (\ref{chieq}). 
An equivalent, and perhaps a simpler way to obtain the closed equation 
for $\bchi$ is to eliminate $A$ and $A_N$ from Eq.~(\ref{chieq}) 
by using Eqs.~(\ref{Aeq}) and (\ref{An}). 

An important fact is that the closed equation for $\bchi$ 
is obtained without using the $({}^0_i)$-component or
the traceless part of the $({}^i_j)$-component of the 
Einstein equations, both of which are absent in the background 
equations. We also see that Eq.~(\ref{An}) exactly corresponds to
the perturbation of Eq.~(\ref{dHdN}). This indicates that we are on
the right track.
If we obtain a complete set of solutions of Eq.~(\ref{chieq})
(supplemented by Eqs.~(\ref{Aeq}) and (\ref{An})), 
the only remaining task is to solve for $\calR$ and $\ksg$.
Let us denote the general solution of Eq.~(\ref{chieq}) by
\begin{equation}
 \bchi=c^\alpha\bchi_{(\alpha)}\,, 
\end{equation}
where $\alpha$ runs from $1$ to $2n$, $\bchi_{(\alpha)}$ are the $2n$ 
independent solutions and $c^\alpha$ are arbitrary constants.

{}From the long wavelength limit of the traceless 
$({}^i_j)$-component, Eq.~(\ref{ijTL}), we have
\begin{equation}
\ksg\propto {1\over a^2}.
\label{ksgeq}
\end{equation}
Under the present gauge condition (\ref{gauge}), this implies
\begin{equation}
 \calR_N={\ksg\over3aH}\propto{1\over a^3 H}\,.
\label{calRN}
\end{equation}
On the other hand, the $({}^0_i)$-component of the perturbed
Einstein equations, Eq.~(\ref{0ieq}), is rewritten as 
\begin{equation}
 \calR_N=A-{1\over 2}\bphi_N\cdot \bchi.
\label{0ieqN}
\end{equation}
Using Eqs.~(\ref{bgN}), (\ref{frN}) and (\ref{Aeq}), 
$\calR_N$ can be expressed in terms of $\bchi$ as 
\begin{eqnarray}
 \calR_N
&=&{H^2\over 2 U}\left({d\bphi_N\over dN}\cdot\bchi
    -\bphi_N\cdot{d\bchi\over dN}\right).
\end{eqnarray}
Hence from Eq.~(\ref{calRN}), we find
\begin{equation}
W[\bchi]:={a^3H^3\over2U}\left({d\bphi_N\over dN}\cdot\bchi
    -\bphi_N\cdot{d\bchi\over dN}\right)
=a^3 H\calR_N
\end{equation}
should be a constant, which can be directly proved
 by using Eq.~(\ref{chieq}) with the aid of Eq.~(\ref{Aeq}). 

Now, we can calculate $\calR$ and $\ksg$
for each independent solution of 
the scalar field perturbation.  We obtain 
\begin{equation}
 \calR_{(\alpha)}=W_{(\alpha)}\int^{N}_{N_b}{dN\over a^3 H}\,,   
\quad \ksg{}_{(\alpha)}={3W_{(\alpha)}\over a^2}\,,
\end{equation}
where $W_{(\alpha)}$ is a constant given by 
\begin{equation}
 W_{(\alpha)}=W[\bchi_{(\alpha)}],
\label{wron}
\end{equation}
and $N_b$ can be arbitrary chosen. We leave it unspecified here.

The arbitrariness of the choice of $N_b$ is the reflection of
a residual gauge degree of freedom in the present gauge.
Note that there is a trivial solution $\bchi=0$ for which
$\calR=c^0={\rm constant}$. The addition of this solution
corresponds to a variation of $N_b$.
Hence we have $2n+1$ integration 
constants for the whole set of the perturbation equations 
in this gauge. This is not a contradiction because 
there is a residual gauge degree of freedom 
in the gauge $B=H'_L=0$. Under an infinitesimal 
transformation of the time coordinate,
\begin{equation}
N\to N-\delta N, 
\end{equation}
$H_L$ transforms as
\begin{equation}
H_L\to H_L+\delta N. 
\end{equation}
Hence the gauge condition $H'_L=0$ allows a further gauge 
transformation given by $\delta N= c=$ constant, which corresponds
to an infinitesimal time translation mode. 
Applying this gauge transformation to 
the null perturbation, we obtain a pure gauge mode,
\begin{eqnarray}
 \bchi =c\,\bphi_N,\quad \calR=c. 
\end{eqnarray}
This fact implies that one of the $2n$ solutions of 
Eq.~(\ref{chieq}) should be proportional to this time translation
mode. In fact, we can verify that 
$\bchi=\bphi_N$ is a solution of Eq.~(\ref{chieq}) 
by direct substitution. 
Hence we may set 
\begin{equation}
\bchi_{(1)}=\bphi_N. 
\end{equation}
Then we also find $W_{(1)}=0$. 

The issue of the number of physical degrees of freedom becomes
transparent by constructing a gauge invariant quantity. A
convenient choice is the scalar field perturbation $\bchi_F$
on the flat hypersurface (defined by $\calR=0$).
It is given by
\begin{equation}
 \bchi_F=\bchi-\bphi_N\calR. 
\label{defchiF}
\end{equation}
Then the general solution for $\bchi_F$ is given by 
\begin{equation}
\bchi_F= c^\alpha\left(\bchi_{(\alpha)}- W_{(\alpha)}\bpsi\right),
\label{chiF}
\end{equation}
where $W_{(1)}=0$ and we have redefined $c^1$ by the replacement; 
$c^1-c^0\to c^1$, and
\begin{equation}
\bpsi:= \bphi_N \int_{N_{b}}^{N}{dN\over e^{3N} H}\,.
\end{equation} 
One sees that the change of $N_b$ is always absorbed in the 
redefinition of $c^1$. 
Hence, only the $2n$ integration constants remain. 

Since there exists at least 
one solution that has a non-vanishing $W$, we may assume 
$W_{(2)}\ne 0$ without any loss of generality. 
Then it is worthwhile to mention that it is always possible to set 
$W_{(\alpha)}=0$ for $\alpha\ne 2$ by redefinition of 
the complete set of solutions.
In fact, $W$ vanishes for the linear combination of the solutions 
defined by $\displaystyle
\bchi_{(\alpha)} - {W_{(\alpha)}\over W_{(2)}}\bchi_{(2)}$ 
for $\alpha\ge 3$.
It may be also useful to note that 
$c^{\alpha}W_{(\alpha)}$ describes the amplitude
of the adiabatic decaying mode\cite{KodHam},
 while the adiabatic growing mode amplitude is given by $c^1$.

Let us now consider the perturbation of the background equations
(\ref{bgN}) and (\ref{frN}). Let us assume that the general solution
 for the background equations is known. 
Except for the trivial time translation, the general 
solution contains $2n-1$ integration constants, $\lambda^a$, 
which distinguish the various solutions. Each solution gives 
a curve in the phase space of the scalar field 
parametrized by some time coordinate $\lambda^1$. 
Let $\hat\chi^A=(\chi^p, \pi^q)$ 
($A=1,2,\cdots,2n$) be the phase space coordinates where $\pi^q$
are the momentum variables.
Then the general solution is expressed
as $\hat\chi^A=\hat\chi^A(\lambda^\beta)$. We may 
regard this as a coordinate transformation of the phase space
coordinates. We see that the perturbation of a given background 
solution is given by the Jacobian 
$\partial\hat\chi^A/\partial\lambda^\beta$, including 
the time translation mode. It is then easy to convince ourselves
that a necessary condition for these background solutions to describe
the solutions of the perturbation in the long wavelength 
limit is the non-disturbance of the time coordinate under the
presence of a perturbation, since the commutable property
of the partial derivatives; 
$\partial^2/\partial\lambda^\alpha\partial\lambda^\beta
=\partial^2/\partial\lambda^\beta\partial\lambda^\alpha$, 
should be maintained.
As we have seen, the $e$-folding number $N=\ln a$
is indeed such a coordinate for the gauge $B=H_L'=0$.
Thus we take $N$ as the time coordinate in the background 
equations and set $\lambda^\alpha=(N,\lambda^a)$.
Note that the background equations (\ref{bgN}) and (\ref{frN})
have no explicit $N$-dependence. Hence $\bphi_N$ is a solution 
to the perturbed background equations, which corresponds to 
the time translation mode.

Taking the derivative of Eqs.~(\ref{bgN}) and (\ref{frN}) 
with respect to $\lambda^{\alpha}$, we obtain
\begin{eqnarray}
&& H{d\over dN} H {d\over dN}\phi_{\lambda}^p
  +3H^2\phi_{\lambda}^p+U^{|p}{}_{|q}\phi_{\lambda}^q
-2U^{|p}{H_{\lambda}\over H} 
+ H^2{\phi_N^p}{d\over dN}{H_{\lambda}\over H}=0,
\label{philameq}\\
 &&2U{H_\lambda\over H}
   ={H^2}\bphi_N\cdot{d\over dN}\bphi_{\lambda}
     + U_{|p} \phi^p_{\lambda},
\label{Hlameq}
\end{eqnarray}
where the suffix $\lambda$ represents 
$\partial/\partial\lambda^{\alpha}$.
We find these equations are equivalent to 
those for the perturbation in the long wavelength limit 
in the $B=H'_L=0$ gauge, Eqs.~(\ref{chieq}) and (\ref{Aeq}),
respectively, with the identifications $\bphi_\lambda=\bchi$
and $H_\lambda/H=-A$.
 Further, as we have already seen, the 
$\lambda^\alpha$-derivative of Eq.~(\ref{dHdN}) is equivalent
to Eq.~(\ref{An}).
Thus we conclude that a complete set of 
solutions in the long wavelength limit, $\bchi_{(\alpha)}$, can be 
constructed from the solutions of the background equations;
$\bchi_{(\alpha)}=\partial\bphi/\partial\lambda^\alpha$. 
Once $\bchi_{(\alpha)}$ are obtained, the corresponding set of
gauge-invariant quantities $\bchi_{F(\alpha)}$ is readily
obtained as
\begin{equation}
\bchi_{F(\alpha)}=\bchi_{(\alpha)}-W_{(\alpha)}\bpsi,
\end{equation}
where $W_{(\alpha)}$ is given by Eq.~(\ref{wron}).

Before closing this subsection, we mention the generalization
of the scalar field metric $h_{pq}$ to a non-trivial one.
Except for the perturbed field equation (\ref{chieq}) or
 (\ref{philameq}),
the only modification is to replace all the derivatives in
the equations with the covariant ones with respect to $h_{pq}$. 
For example,
\begin{equation}
{d\over dN}\phi_\lambda^p\to
{D\over dN}\phi_\lambda^p={d\over dN}\phi_\lambda^p
+\Gamma^p{}_{qr}\phi_\lambda^q\phi_N^r\,,
\end{equation}
where $\Gamma^p{}_{qr}$ is the connection of $h_{pq}$.
As for the perturbed field equation, we have to add the curvature
term in addition to the covariantization of the derivatives.
That is, Eq.~(\ref{chieq}) is modified as
\begin{equation}
  H{D\over dN}\left(H \chi^p_N\right)
+3H^2\chi^p_N +H^2R^p{}_{rqs}\phi_N^r\phi_N^s\chi^q
 +U^{|p}{}_{|q}\chi^q
  +2U^{|p} A - H^2{\phi^p_N} A_N =0, 
\label{covchieq}
\end{equation}
and similarly for Eq.~(\ref{philameq}).

\subsection{Curvature perturbation on the comoving hypersurface}

Among the various geometrical quantities, one of the most convenient
representations of the perturbation amplitude is the curvature
perturbation of the comoving hypersurface, $\calR_c$. 
For the adiabatic growing mode perturbation, 
it is known to stay constant on super-horizon scales\cite{KodSas}.
Here, we relate $\calR_c$ with the scalar field perturbation and present
a formula that can be used to evaluate $\calR_c$ at the end of inflation
in terms of the initial data of the scalar field perturbation. Then we
take the slow rolling limit and show how our formula reduces
to the one obtained in Ref.~\citen{SasSte}. To obtain $\calR_c$ that is
directly relevant to observational quantities, one must solve the
evolution of $\calR_c$ during the reheating stage after inflation.
One may also have to evaluate the possible contribution of the
isocurvature perturbation. We defer these issues to future studies
and derive here a formula valid up to a time $N=N_e$ before the
reheating commences. It may be noted, however, that there are situations
in which one can neglect the time variation of $\calR_c$ during
reheating. For example, if all the trajectories in the phase space of
the scalar field converge to one path before reheating, the evolution
during reheating is essentially the same as the case of single-scalar
inflation and there will be no additional change
in the amplitude of $\calR_c$.

As we have seen in the previous section, the spacetime configuration of
the scalar field on super-horizon scales is completely determined by 
the background solutions and the parameters $\lambda^\alpha$
characterizing the background solutions play the role of phase 
space coordinates. 
This implies the value of the scalar field
depends on the spacetime coordinates $\{x^\mu\}$ only through
the phase space coordinates for the spatially homogeneous configurations; 
$\bphi(x^\mu)=\bphi\left(\lambda^\alpha(x^\mu)\right)$.
Thus it will be useful to express $\calR_c$ in the language of
the phase space of the background scalar field.

The comoving hypersurface is determined by the condition, 
$T^0_i=0$, 
i.e.,  
\begin{equation}
 \bphi_N\cdot d\bphi=0, 
\label{comoving}
\end{equation}
In general, we cannot expect this condition to determine a 
surface in the phase space of the scalar field since it
is not integrable.
However, in the case of linear perturbation, this condition 
reduces to 
\begin{equation}
\bar\bphi_N(N)\cdot(\bphi-\bar\bphi(N))=0, 
\label{comoving2}
\end{equation}
where the barred quantities represent the background values.
Now it is manifest that this condition determines a surface 
in the phase space. For each $\lambda^a$, the value of $N$ at 
which the comoving surface in the phase space is crossed is different.
We denote this difference by $\Delta N$.
Namely,
\begin{equation}
\bar\bphi_N(N)\cdot\left(\bphi(N+\Delta N,\lambda^a)
-\bar\bphi(N)\right)=0, 
\label{comoving3}
\end{equation}
where $\Delta N$ is a function of $\lambda^\alpha=(N,\lambda^a)$. 
{}From the spacetime point of view, through the dependence of
$\lambda^\alpha$ on $x^\mu$, $\Delta N$ is a function of $x^\mu$. 
Hence we may consider the infinitesimal coordinate transformation
$N\to N-\Delta N$ and move from the $H_L'=B=0$ gauge to the comoving
gauge. Then we have
\begin{equation}
\calR_c(N)=\calR(N)+\Delta N\,.
\end{equation}
On the other hand, the comoving condition (\ref{comoving3})
is rewritten as
\begin{equation}
\bphi_N(N)\cdot(\bchi(N)+\bphi_N(N)\Delta N)=0.
\end{equation}
Thus we have $\Delta N=-(\bphi_N\cdot\bchi)/\bphi_N^2$.
Therefore we obtain
\begin{equation}
\calR_c=-c^{\alpha}
    {\bphi_N\cdot\bchi_{(\alpha)}\over \bphi_N^2} 
    + c^{\alpha}W_{(\alpha)}\int_{N_b}^N{dN\over e^{3N} H}\,, 
\label{calRcform}
\end{equation}
where we remind that 
$\bchi_{(\alpha)}=\partial\bphi/\partial\lambda^\alpha$.
Given the background solutions, this gives the evolution of
$\calR_c$ up to $N=N_e$.

In the above, we have derived the expression for $\calR_c$
in the language of the phase space for the background
solutions. But, of course, it can be derived in the language
of the standard linear perturbation theory.
In the defining equation for $\bchi_F$, Eq.~(\ref{defchiF}),
 we evaluate the right hand side
on the comoving hypersurface to obtain
\begin{equation}
\bchi_F=\bchi_c-\bphi_N\calR_c\,,
\end{equation}
where $\bchi_c$ is the scalar field perturbation on the comoving
hypersurface. Then taking the inner product of the above equation 
with $\bphi_N$ and using the comoving hypersurface condition 
$\bphi_N\cdot\bchi_c=0$, we find
\begin{equation}
\calR_c=-{\bphi_N\cdot\bchi_F\over\bphi_N^2}\,.
\end{equation}
Inserting the expression (\ref{chiF}) for $\bchi_F$ to 
this equation gives Eq.~(\ref{calRcform}).

The remaining task is to determine $c^{\alpha}$. 
In the inflationary universe scenario, scalar field perturbations
are generally induced by quantum vacuum fluctuations.
The evaluation of the vacuum fluctuations is most conveniently done
with respect to $\bchi_F$ since there exists a closed action for
$\bchi_F$ that gives a complete description of the scalar-type 
perturbation\cite{Mukhanov}. Hence we assume that $\bchi_F$ is given at
some initial epoch when the wavelength exceeds the horizon scale.

Suppose that the initial condition of $\bchi_F$ and $\displaystyle 
{D\bchi_F\over dN}$ is given at a time, $N=N_0$. 
The constant $c^{\alpha}W_{(\alpha)}$ can be readily evaluated
by inserting $\bchi_F$ into the operator $W$:
\begin{equation}
W[\bchi_F]=c^{\alpha}W_{(\alpha)}
\left(1+{H^2\over2U}\bphi_N^2\right)
={3H^2\over U}c^{\alpha}W_{(\alpha)}\,.
\end{equation}
Hence,
\begin{equation}
c^{\alpha}W_{(\alpha)}=\left[{a^3H\over6}
\left({D\bphi_N\over dN}\cdot\bchi_F
 -\bphi_N\cdot{D\bchi_F\over dN}\right)\right]_{N_0}\,.
\label{cW}
\end{equation}
Then adopting the notation $\hat\chi^{A}:=\chi^{A}$ 
for $A\leq n$ and $\hat\chi^{A}:=\chi_N^{A-n}$ 
for $A>n$, $c^{\alpha}$ is evaluated from Eq.~(\ref{chiF}) as 
\begin{equation}
 c^{\alpha}=\left[(\hat\chi^{-1})^{(\alpha)}_{A}
 \left\{\hat\chi_{F}^{A}+(c^{\beta}W_{(\beta)})\hat\psi^A\right\}
\right]_{N_0}, 
\label{calpha}
\end{equation}
where $\hat\bchi_F$ and $\hat\bpsi$ are defined in the same 
way as $\hat\chi$, and $(\hat\chi^{-1})^{(\alpha)}_A$ is
the inverse matrix of $\hat\chi^A_{(\alpha)}$, i.e.,
\begin{equation}
(\hat\chi^{-1})^{(\alpha)}_A
={\partial\lambda^\alpha\over\partial\hat\chi^A}\,.
\label{invJ}
\end{equation}

So far our discussion has been completely general. Now we consider
the slow rolling limit. We assume all the components of the
scalar field satisfy the slow rolling condition:
\begin{equation}
\bphi_N^2\ll 1,\quad \left|{D\bphi_N\over dN}\right|\ll |\bphi_N|.
\label{srcond}
\end{equation}
In this limit, the background equations reduce to
\begin{equation}
\phi_N^p=-{U^{|p}\over3H^2}=-(\ln U)^{|p}\,,
\quad H^2={1\over3}U\,.
\label{srbg}
\end{equation}
Thus the momentum variables cease to be dynamical and
the $2n$-dimensional phase space reduces to the $n$-dimensional
configuration space of the scalar field.
Correspondingly the number of the parameters $\lambda^\alpha$
reduces to $n$. 
The perturbed field equation is very simple:
\begin{equation}
{D\over dN}\phi_\lambda^p
=-(\ln U)^{|p}{}_{|q}\phi_\lambda^q\,,
\label{srfeq}
\end{equation}
where we have assumed the scalar space curvature is small;
$|R^p{}_{rqs}|\ll1$.\footnote{It seems this condition is necessary
for the slow rolling approximation to be consistent. However, 
we are unable to give a proof.}
{}From this equation, we easily see that $W[\bphi_\lambda]=0$
 for all the $n$-independent solutions.
Thus the slow rolling assumption kills the adiabatic decaying mode
as in the case of single-scalar inflation.
This implies that $\bchi_F$ is directly given by the background
solutions:
\begin{equation}
\bchi_F=c^\alpha\,\bchi_{(\alpha)}
=c^{\alpha}\,{\partial\bphi\over\partial\lambda^\alpha}\,,
\end{equation}
where $\alpha$ now runs from $1$ to $n$. 
In other words, the flat slicing does not perturb 
the $e$-folding number $N$ in the slow rolling limit.

The formula (\ref{calpha}) for $c^\alpha$ reduces to
\begin{equation}
 c^{\alpha}
=\left[(\chi^{-1})^{(\alpha)}_{p}\,\chi_{F}^{p}\right]_{N_0}
=\left[{\partial\lambda^\alpha\over\partial\phi^p}
\chi_F^p\right]_{N_0}\,. 
\label{srcalpha}
\end{equation}
For definiteness,
we set $\bchi_{(1)}=\bphi_N$ (i.e., $\lambda^1=N$) as before.
The curvature perturbation $\calR_c$ is then given by
\begin{eqnarray}
\calR_c&=&-c^\alpha\,{\bphi_N\cdot\bchi_{(\alpha)}\over\bphi_N^2}
\nonumber\\
&=&-\left\{
\left[{\partial N\over\partial\phi^p}\,\chi_F^p\right]_{N_0}
+\left[{\partial\lambda^a\over\partial\phi^p}\,\chi_F^p\right]_{N_0}
{\bphi_N\cdot\bphi_{,a}\over\bphi_N^2}\right\},
\label{srcalRc}
\end{eqnarray}
where $a=2,3,\cdots,n$ and 
$\bphi_{,a}={\partial\bphi/\partial\lambda^a}$.

If we synchronize the time to be $N=N_e$ at the end of inflation
for all the background scalar field trajectories, we have
$\partial U/\partial\lambda^a=0$ at $N=N_e$. 
Then using the slow rolling equation of motion (\ref{srbg}) 
we find
\begin{equation}
0={\partial U\over \partial\lambda^a}
={\partial U\over\partial\phi^p}
{\partial\phi^p\over\partial\lambda^a}
=-U\bphi_N\cdot\bphi_{,a}\,,
\end{equation}
at $N=N_e$. Thus the evaluation of Eq.~(\ref{srcalRc}) at
$N=N_e$ gives
\begin{equation}
\calR_c(N_e)=-
\left[{\partial N\over\partial\phi^p}\,\chi_F^p\right]_{N_0}\,.
\end{equation}
This formula coincides with the one derived by Sasaki and
Stewart\cite{SasSte}.\footnote{In Ref.~\citen{SasSte}, 
the minus sign on the right hand side of this formula is absent. 
The reason is that $N$ there is defined
as the $e$-folding number counted backwards from the time $N=N_e$
for convenience of the spectrum evaluation.}

\section{Quasi-Nonlinear Perturbation}

An advantage of the formalism developed in the previous section
is that it can be extended so as to take into account the 
non-linearity of the scalar field dynamics. 
The basic idea is to consider the linearization with respect 
only to inhomogeneities of the spatial metric. 
For simplicity, we again assume the trivial metric for the scalar 
field space, $h_{pq}=\delta_{pq}$.

We consider the following metric form,
\begin{equation}
 ds^2=-{1\over \tilde H^2} dN^2+ e^{2N} 
  \bigl((1+2H_L)\delta_{ij}+2H_{Tij}\bigr) dx^i dx^j, 
\end{equation} 
where $\tilde H$ depends both on $N$ and $x^i$. Here and
in what follows the tilde is attached to a non-linearly perturbed
 quantity.
We consider $H_L$ and $H_{Tij}$ to be small of $O(\epsilon)$.
In accordance with the linear case, we assume the existence
of a potential function $H_T$ for $H_{Tij}$ such that
\begin{equation}
-\nabla^2H_{Tij}
=\left[\nabla_i\nabla_j-{1\over3}\delta_{ij}\nabla^2\right] H_T\,.
\end{equation}
We also assume the spatial derivatives of
 $\tilde H$ and $\tilde\bphi$ are small of $O(\epsilon)$.
This assumption corresponds to taking the long wavelength limit.
Furthermore, we assume the gauge condition $dH_L/dN=O(\epsilon^2)$
can be imposed consistently.
We then linearize the scalar field and Einstein equations
with respect to $\epsilon$ while keeping the other nonlinear terms
as they are.

First let us consider the scalar field equation.
Keeping the terms linear in $\epsilon$, 
the field equation reduces to 
\begin{eqnarray}
0 & = & {1\over \sqrt{-g}}\partial_{\mu}\left[ g^{\mu\nu}\sqrt{-g}
\partial_{\nu}\tilde \phi^p\right]
\cr
&=&
{\tilde H\over e^{3N}} (1-3H_L) {d\over dN} \tilde H e^{3N} 
 (1+3 H_L){d\over dN} {\tilde\phi}^p+U(\tilde\bphi)^{|p}
\cr
 &=& {\tilde H} {d\over dN} \tilde H 
 {d\over dN} {\tilde \phi}^p+3 \tilde H^2{d\over dN}\tilde\phi^p 
 +U(\tilde \bphi)^{|p}. 
\label{fieldeq}
\end{eqnarray}
This is the same form as the background equation (\ref{bgN}). 

Next let us consider the $({}^0_0)$-component of the Einstein
 equations. To the linear order in $\epsilon$, we have 
\begin{eqnarray}
&&G^0{}_0=-3 \tilde H^2, 
\cr
&&T^0{}_0=-{1\over 2} \tilde H^2 \tilde\bphi_N^2 -U(\tilde\bphi).
\end{eqnarray}
Therefore we find
\begin{equation}
 \tilde H^2\left(1-{1\over 6}\tilde\bphi_N^2\right)
 ={1\over 3}U(\tilde\bphi), 
\label{friedmann2}
\end{equation}
which is also the same form as the background equation (\ref{frN}).

Thus we find that, in the long wavelength limit,
solving the equations for the perturbed scalar field
in the $dH_L/dN=0$ gauge is completely equivalent to
finding the spatially homogeneous background solutions. 

The remaining task is to determine $\calR:=H_L+{1\over3}H_T$ 
and $\ksg:=e^N\tilde H\,dH_T/dN$ as before
(note that $\ksg$ is not $k$ times $\sigma_g$ but
regarded as a single symbol here).
The traceless part of the $({}^i_j)$-component of the 
Einstein equations is essentially the same as
Eq.~(\ref{ijTL}) also in the present case. 
This tells us that we have 
\begin{equation}
k\sigma_g=3e^N\tilde H\calR_N\propto {1\over e^{2N}}\,. 
\label{Rp1}
\end{equation}
The $({}^0_i)$-component of the Einstein tensor becomes 
\begin{equation}
 G^0{}_i=\nabla_i \tilde H^2+2\tilde H^2\nabla_i \calR_N\,,
\end{equation}
while that of the energy momentum tensor is given by 
\begin{equation}
 T^0{}_i= -{\tilde H^2} \tilde \bphi_N \cdot 
    \nabla_i\tilde \bphi\,. 
\end{equation}
Hence we have 
\begin{equation}
 e^{3N} \tilde H\nabla_i \calR_N 
=-e^{3N}\left(\nabla_i \tilde H
 +{1\over2}{\tilde H}\,\tilde\bphi_N \cdot\nabla_i\tilde\bphi\right), 
\label{dcalRN}
\end{equation}
which is expected to be constant in time because of
Eq.~(\ref{Rp1}). In fact, we find 
\begin{eqnarray}
 {d\over dN}(e^{3N}\tilde H\nabla_i \calR_N)
&=&-{d\over dN}
\left[e^{3N}\left(\nabla_i \tilde H
 +{1\over2}{\tilde H}\,\tilde\bphi_N \cdot\nabla_i\tilde\bphi
\right)\right]
\nonumber\\
&=&{e^{3N}\over 2 \tilde H}\left(
   -{2U(\tilde\bphi)\over \tilde H}
    \nabla_i \tilde H + \tilde H^2 \tilde\bphi_N
    \nabla_i \tilde\bphi_N 
    +\nabla_i U(\tilde\bphi)\right)
\nonumber\\
&=&0,
\end{eqnarray}
where Eqs.~(\ref{fieldeq}) and (\ref{friedmann2}) are used
in the second equality and the spatial derivative 
of Eq.~(\ref{friedmann2}) in the last equality.
This verifies the consistency of our assumptions, in particular
the gauge condition $dH_L/dN=O(\epsilon^2)$.

Let us reinterpret the discussions given in \S\S 2.3
in the present context. For simplicity we assume that the 
linear perturbation is valid at $N=N_0$. Otherwise, the evaluation of
the initial perturbation becomes too difficult. 
As in \S\S 2.3, we give the initial data in terms of $\bchi_F$ 
and $d\bchi_F/dN$.

Then from Eq.~(\ref{Rp1}), the evolution of $\calR$ is given by 
\begin{equation}
 \calR=C_{\calR} \int_{N_0}^N {dN\over e^{3N} \tilde H}, 
\end{equation}
where the coefficient $C_{\calR}$ is expressed in terms of the initial
values of $\bchi_F$ and $d\bchi_F/dN$ as
\begin{equation}
 C_{\calR}=\left[{e^{3N}\tilde H\over6}
\left({d\bphi_N\over dN}\cdot\bchi_F
  -\bphi_N\cdot{d\bchi_F\over dN}\right)\right]_{N_0}\,,
\end{equation}
which corresponds to Eq.~(\ref{cW}) for the linear case.
In the above, we have chosen $\calR(N_0)=0$ by setting $N_b=N_0$ 
for convenience.

Using Eq.~(\ref{chiF}), the initial data for the scalar field are given
by 
\begin{eqnarray}
&& \tilde\bphi(N_0)=\bphi(N_0)+\bchi_F(N_0),
\nonumber\\
&& \tilde\bphi_N(N_0)
=\left(1-{C_{\calR}\over e^{3N_0}\tilde H_0}\right)\bphi_N(N_0)
+\bchi_{F\,N}(N_0)\,,
\end{eqnarray}
where $\tilde H_0=\tilde H(N_0)$ and $\bchi_{F\,N}=d\bchi_F/dN$.
The evolution of $\tilde\bphi$ is determined by solving 
the nonlinear background field equation (\ref{fieldeq})
supplemented with Eq.~(\ref{friedmann2}).

In \S\S 2.3, to evaluate the amplitude of the perturbation at the end
of inflation, it was necessary to move to the comoving hypersurface.
In terms of the background solutions,
this amounts to finding a comoving surface in the phase 
space. However, as noted there, the comoving condition
(\ref{comoving}) does not specify a surface in the phase space
in the nonlinear case. To circumvent this difficulty,
here we propose to use the constant Hubble hypersurface 
in substitution of the comoving one.
It is apparent from Eq.~(\ref{friedmann2}) that there exists a 
surface in the phase space corresponding to the constant 
Hubble hypersurface.
{}From the fact that the right hand side of Eq.~(\ref{dcalRN}) is 
constant in time, we find 
\begin{equation}
 \nabla_i \ln \tilde H
    =-{1\over 2}\tilde\bphi_N \cdot \nabla_i 
     \tilde\bphi+{d_i\over e^{3N}\tilde H}\,, 
\label{dlogH}
\end{equation}
where $d_i$ is a time-independent vector. 
At a later epoch, the second term in the right hand side 
can be neglected and the difference between the comoving 
gauge and the constant Hubble gauge becomes negligibly small.  

Then as a function of $\bchi_F(N_0)$ and $\bchi_{F\,N}(N_0)$, 
we define $\Delta N_H$ as the 
difference of the time to cross the constant Hubble 
surface in the phase space, namely, by the condition,
\begin{equation}
\tilde H\bigl(\bphi(N+\Delta N_H),\bphi_N(N+\Delta N_H)\bigr)
=\hbox{independent of $\lambda^a$}.
\end{equation}
Finally, we find the curvature perturbation on the 
constant Hubble hypersurface $\calR_H$ as 
\begin{equation}
 \calR_c\simeq\calR_H=\calR+ \Delta N_H\,.  
\label{nlcalRc}
\end{equation}

In the case the slow rolling condition (\ref{srcond}) is satisfied
at the initial time, the first term in the right hand side of 
Eq.~(\ref{nlcalRc}) can be totally neglected and we simply have 
\begin{equation}
 \calR_c\simeq\calR_H= \Delta N_H\,, 
\end{equation}
where $\Delta N_H$ will now be a function of $\bchi_F(N_0)$ alone.

\section{Summary}

In this paper, we have investigated the dynamics of a multi-component
scalar field on super-horizon scales, i.e., in the long wavelength
limit. We have shown that there is a simple relation between 
the perturbation equations in the long wavelength limit and the
background equations. That is, the derivative of the general
solution of the background equations with respect to a parameter
that characterizes different solutions satisfies the same equation
as the perturbation in the long wavelength limit does.
However, we have also found that the explicit 
form of the relation depends on the choice of gauge, and the
choice of gauge corresponds to that of a time coordinate in the 
phase space of the background scalar field.
 We have found that the simplest form of the relation
is obtained in the gauge in which the 
$e$-folding number $N$ of cosmic expansion is 
unperturbed even under the presence of the perturbation. 
Then using this result, we have given a method
to calculate the amplitude of the spatial curvature perturbation
on the comoving hypersurface, $\calR_c$, from the knowledge of 
the background solutions alone. 

As a natural extension of our approach, we have considered to take 
into account the nonlinearity of the scalar field dynamics
in the perturbation. We have found that this can be actually done 
under several reasonable assumptions. The result provides a 
powerful tool to evaluate the effect of nonlinearity of the scalar
field potential during the inflationary stage. 
In particular, the effect of non-gaussian statistics of the 
perturbation, given a gaussian distribution of the initial perturbation
due to quantum vacuum fluctuations, can be evaluated by
studying the background solutions alone.

The present nonlinear extension is similar to the 
so-called anti-Newtonian approximation\cite{Tomita} or 
the gradient expansion method\cite{gradexp}. It may be worthwhile
to clarify how our result is related to these methods.

\section*{Acknowledgements}
We would like to thank T. Hamazaki, H. Kodama, Y. Nambu and
A. Taruya for useful discussions. This work was supported
by the Monbusho Grant-in-Aid for Scientific Research, No.~09640355.

\appendix
\section{Other Choices of Gauge}

In this appendix, we compare the 
long wavelength perturbation equations and the background equations
in a couple of other choices of gauge. We find the choice of the
time coordinate in the scalar field phase space determines the
corresponding gauge condition for the perturbation as in the
case of the $H_L'=B=0$ gauge. However, the resulting
equations turn out to be less tractable.

\subsection{$t$ as a time coordinate: the synchronous gauge}

In the synchronous gauge, $A=B=0$, the cosmological time
$t$ remains to be the proper time along curves normal to the
constant time hypersurfaces. Hence we expect $t$ to be the 
relevant time coordinate to parametrize the background solutions.

The perturbation equations (\ref{feqpert}) and (\ref{00eq}) 
in the long wavelength limit in the synchronous gauge become
\begin{eqnarray}
 &&{\ddot\chi^p}+3H {\dot\chi^p}
 -{\dot\phi^p}\left({\ksg\over a} -3\dot\calR\right)
 +U^{|p}{}_{|q}\chi^q=0.
\label{chieqsyn}
\\
 && 2H\left(3\dot\calR-{\ksg\over a}\right)
=\dot\bphi\cdot\dot\bchi+U_{|p}\chi^p\,.
\label{Aeqsyn}
\end{eqnarray}
Eliminating $\displaystyle\left(3\dot\calR-{\ksg\over a}\right)$ 
from these equations, we obtain an
equation written in terms of $\bchi$ alone. 
Note that $\displaystyle\left(\dot\calR-{\ksg\over3a}\right)$
in this gauge represents 
the perturbation of the Hubble parameter
as seen from Eq.~(\ref{deftilH}).

Once we know the solutions for $\bchi$, 
we can evaluate the spatial curvature perturbation 
with the aid of the $({}^0_i)$-component of the Einstein
equations, Eq.~(\ref{0ieq}), which now reduces to 
\begin{equation}
 \dot\calR= -{\dot\bphi\cdot \bchi\over 2}.
\label{dotcalR}
\end{equation}
Integrating this equation, $\calR$ can be evaluated. 
But different from the $H_L'=B=0$ gauge, we have no knowledge
 of the explicit time dependence of the integrand at all. 
So the integration must be done for each mode 
separately. 

On the other hand, the time dependence of $\ksg$ can be found
from the traceless part of the $({}^i_j)$-component,
Eq.~(\ref{ijTL}), as
\begin{equation}
\ksg\propto {1\over a^2}\,.
\end{equation}
Using Eqs.~(\ref{Aeqsyn}) and (\ref{dotcalR}), this implies
\begin{equation}
\ksg a^2={a^3\over2H}(\ddot\bphi\cdot\bchi
-\dot\bphi\cdot\dot\bchi)=\hbox{constant}\,.
\end{equation}

As before, the general solution of Eq.~(\ref{chieqsyn}) with
(\ref{Aeqsyn}) contains $2n$ integration constants. 
When we evaluate $\calR$, there appears an additional 
integration constant. Hence, we have $2n+1$ integration 
constants in total. This situation is exactly parallel to
the case of the $H_L'=B=0$ gauge.
Namely, one mode is responsible for a
gauge degree of freedom, since the synchronous gauge condition
 allows an additional gauge 
transformation given by $t\to t+c$, where $c$ is 
a constant. 
Then the gauge mode is found to be 
\begin{eqnarray}
 \bchi = c\,{\dot\bphi},\quad \calR=c\,H.
\end{eqnarray}
Also parallel to the $H_L'=B=0$ gauge, there exists
a trivial solution:
\begin{eqnarray}
 \bchi = 0,\quad \calR=\hbox{constant}.
\end{eqnarray}
As before, this gives one of the solutions for the gauge-invariant
scalar field perturbation $\bchi_F$ as
\begin{equation}
 \bchi_{F}=\bchi-{\dot\bphi\over H}\calR\propto\bphi_{N}\,. 
\end{equation}

Now let us turn to the comparison of the perturbation equations
with the background equations. As we have mentioned in the 
beginning, since the cosmological time is unperturbed in 
the synchronous gauge, the relevant time coordinate will be $t$.
Thus taking the derivatives of Eqs.~(\ref{bgcos}) and (\ref{frcos})
with respect to $\lambda^{\alpha}=(t,\lambda^{a})$, 
we have 
\begin{equation}
 \ddot\phi_{\lambda}^{p}+3H_{\lambda} 
\dot\phi^{p} +3H \dot\phi^{p}_{\lambda} + 
 U^{|p}{}_{|q}\phi^{q}_{\lambda}=0\,, 
\end{equation}
and 
\begin{equation}
 2HH_{\lambda} ={1\over 3}
 \left(\dot\bphi\cdot\dot\bphi_{\lambda}
  + U_{|p}\phi^{p}_{\lambda}\right), 
\end{equation}
where the suffix $\lambda$ again represents the partial 
derivative with respect to $\lambda^{\alpha}$. 
It is readily seen that 
these are equivalent to Eqs.~(\ref{chieqsyn}) and (\ref{Aeqsyn}), 
with the identifications, 
\begin{equation}
\bphi_\lambda=\bchi,\quad 
H_{\lambda}=\dot\calR-{1\over3}{\ksg\over a}\,. 
\end{equation}

\subsection{$H$ as a time coordinate: the constant Hubble gauge}

As another example, we consider to take the Hubble parameter $H$
as a time coordinate. Then the relevant choice of gauge will be
to take the constant Hubble gauge in which $H$ is unperturbed 
under the presence of the perturbation.

{}From the expression for the perturbed Hubble parameter,
Eq.~(\ref{deftilH}), we find the condition for the
constant Hubble gauge as
\begin{equation}
\calh A-\calR'+{1\over 3} \ksg=0. 
\end{equation}
In this gauge, 
the perturbation equations (\ref{feqpert}) and (\ref{00eq}), 
in the long wavelength limit, reduce to 
\begin{eqnarray}
 &&{\chi^p}''+2\calh {\chi^p}'-\left\{
    2{\phi^p}''+ \calh {\phi^p}'\right\} A -{\phi^p}' A'
     +a^2 U^{|p}{}_{|q}\chi^q=0,
\label{chieqcom}\\
 && A 
={1\over {\bphi'}^2}\left(\bphi'\cdot\bchi'
     +a^2 U_{|p}\chi^p\right).
\label{Aeqcom}
\end{eqnarray}

{}From Eq.~(\ref{ijTL}), we find $\ksg\propto 1/a^2$, 
and from Eq.~(\ref{0ieq}), we have 
$-(2/3)\ksg=\bphi'\cdot\bchi$. 
Hence, we find 
\begin{equation}
\ksg a^2 =-{3a^2\over2}\bphi'\cdot\bchi=\hbox{constant}.
\end{equation} 
Taking the time derivative of this equation, 
we find the constraint,
\begin{equation}
 \bphi'\cdot\bchi'=a^2 U_{|p} \chi^{p}\,.
\label{constraint}
\end{equation}
This means that the initial condition for 
the scalar field perturbation cannot be chosen arbitrarily.  
This is because the constant Hubble gauge condition 
completely fixes the time slicing and hence the constant 
time surfaces in the phase space of the scalar field. 
So the perturbation in the direction normal to the
constant time surfaces is not allowed. 

Hence, in the present gauge, the general solution for 
scalar field perturbations has $2n-1$ integration constants. 
The curvature perturbation can be evaluated by 
integrating 
\begin{equation}
 \calR'=\calh A-{1\over 2}\bphi'\cdot\bchi 
   = 2\calh {\bphi'\cdot\bchi'\over {\bphi'}^2} 
          -{1\over 2}\bphi'\cdot\bchi. 
\end{equation}
Again, the integration of this equation is non-trivial.

As in the previous two cases, we have a trivial solution 
\begin{eqnarray}
 \bchi=0,\quad \calR=\hbox{constant}.
\end{eqnarray}
This again tells us that 
$\bphi_N$ is a solution for $\bchi_F$. 

Now let us consider the background equations. 
Using $H$ as the time coordinate,
the background equations are written as 
\begin{eqnarray}
 &&4{1\over \bphi_H^2}{d\over dH} {\phi^p_H\over \bphi_H^2}
 -6{H\over \bphi_H^2}\phi^p_H+ U^{|p}=0, 
\cr
 && H^2={1\over 3}
  \left({2\over \bphi_H^2}+U\right), 
\label{constHbg}
\end{eqnarray}
where subscript $H$ represents the derivative with respect to $H$. 
The latter equation is the equation that constrains 
the scalar field in its phase space. Different from the 
previous two cases, these equations do not have the 
invariance with respect to the time translation because they 
contain $H$ explicitly.
Hence, $\bphi_H$ is not a solution of the perturbation equations. 

Taking the derivative of Eqs.~(\ref{constHbg}) 
with respect to $\lambda=\lambda^{a}$, we obtain 
\begin{eqnarray}
 &&{\phi^p_{\lambda}}''+2\calh {\phi^p_{\lambda}}'-\left\{
    2{\phi^p}''+ \calh {\phi^p}'\right\} 
    {2\bphi'\cdot\bphi'_{\lambda}\over {\bphi'}^2} -{\phi^p}' 
     \left({2\bphi'\cdot\bphi'_{\lambda}\over {\bphi'}^2}\right)'
     +a^2 U^{|p}{}_{|q}\phi_{\lambda}^q=0,
\cr
 && \bphi'\cdot\bphi_{\lambda}'=a^2 U_{|p}\phi_{\lambda}^p.
\end{eqnarray}
These are equivalent to Eqs.~(\ref{chieqcom}) and (\ref{Aeqcom})
supplemented by the constraint (\ref{constraint}).

\end{document}